%% file: rsm7.tex
\documentclass[a4paper,12pt]{article}
\usepackage[dvips]{graphicx}

\newcommand{\be}{\begin{equation}}
\newcommand{\ee}{\end{equation}}
\newcommand{\bea}{\begin{eqnarray}}
\newcommand{\eea}{\end{eqnarray}}
\newcommand{\beaa}{\begin{eqnarray*}}
\newcommand{\eeaa}{\end{eqnarray*}}
\newcommand{\ben}{\begin{enumerate}}
\newcommand{\een}{\end{enumerate}}
\newcommand{\bi}{\begin{itemize}}
\newcommand{\ei}{\end{itemize}}

\def\zzz{\mathop{\rm Z\kern -0.25em Z}\nolimits}

\def\lm{\lambda}

\def\eps{\epsilon}

\def\phi{\varphi}

\def\om{\omega}

\def\Gm{\Gamma}

\def\squarebox#1{\hbox to #1{\hfill\vbox to #1{\vfill}}}

\input graph.tex

\begin{document}

\begin{center}
Bifurcation diagram and pattern formation in superconducting wires
with electric currents
\\[.3cm]
J. Rubinstein \footnote{Mathematics Department, Indiana
University, Bloomington IN 47405}, P.
Sternberg\footnote{Mathematics Department, Indiana University,
Bloomington IN 47405}, and Q. Ma \footnote{Mathematics Department, Indiana
University, Bloomington, IN 47405}\\[.3cm]

\vspace{.2cm}

\begin{minipage}[t]{142.5mm}
\textbf{Abstract. } \textsl { } We examine the behavior of a
one-dimensional superconducting wire exposed to an applied
electric current. We use the time-dependent Ginzburg-Landau model
to describe the system and retain temperature and applied current
as parameters. Through a combination of spectral analysis,
asymptotics and canonical numerical computation, we divide this
two-dimensional parameter space into a number of regions. In some
of them only the normal state or a stationary state or an
oscillatory state are stable, while in some of them two states are
stable. One of the most interesting features of the analysis is
the evident collision of real eigenvalues of the associated
PT-symmetric linearization, leading as it does to the emergence of
complex elements of the spectrum. In particular this provides an
explanation to the emergence of a stable oscillatory state. We
show that part of the bifurcation diagram and many of the emerging
patterns are directly controlled by this spectrum, while other
patterns arise due to nonlinear interaction of the leading
eigenfunctions.
\end{minipage}

\end{center}
\vspace{.5cm} PACS numbers: 74.20.de 74.25.sv 74.25.dw

We consider a one-dimensional superconducting wire of finite
extent. An electric current is fed into the wire at one of its
ends creating a voltage difference across the wire. This is a
canonical problem that has received considerable attention since
it involves the case of a resistive state in which a normal
current and a superconducting current coexist. One of the
intriguing phenomena associated with this state is the formation
of phase slip centers (PSC). These are points in space-time where
the order parameter in the time-dependent Ginzburg Landau equation
(TDGL) vanishes. In fact, as was pointed out by Ivlev and Kopnin
\cite{ivko}, phase slip centers can be thought of as vortices in
space-time. The appearance of phase slip centers is related to
oscillations found numerically through the emergence of
time-periodic solutions. The phase slip centers and the associated
oscillations can be indirectly observed experimentally via the
appearance of steps in I-V curves (\cite{krba}, \cite{krwa},
\cite{vodo}). Another type of behavior found in the resistive
state involves stationary solutions of the TDGL \cite{krba}. In
this case the gauge invariant quantities reach a steady state.

One goal of this paper is to understand the origin of the
different patterns observed in the resistive state. Another goal
is to compute the bifurcation diagram in the parameter space. In
particular we will explain why and when oscillatory solutions
emerge. We will also consider the loss of stability of these
oscillatory solutions as the temperature is lowered below a
critical value that depends on the applied current $I$. The key
idea is that the oscillations appear as a consequence of a Hopf
bifurcation driven by a PT-symmetric spectral problem. A crucial
role in the analysis is played by the dependence of this spectrum
on the applied current. An additional goal is to elucidate the
appearance of hysteresis in I-V curves in the present setup.

Our starting point is the time-dependent Ginzburg Landau model that
we write in a nondimensional form:
\begin{equation}
\psi_t+i\phi \psi=\psi_{xx}+\Gm \psi - |\psi|^2 \psi. \label{g1}
\end{equation}
Here $\psi$ is the complex-valued order parameter, $\phi$ is the
electric potential and $\Gm$ is proportional to $T_c-T$.
Conservation of the current $I$ implies the relation
\begin{equation}
\frac{i}{2}\left(\psi \psi_x^*-\psi_x \psi^*\right) - \sigma
\phi_x=I, \label{g3}
\end{equation}
where $\sigma$ models the Ohmic resistivity. (In equations
(\ref{g1}), (\ref{g3}) and all subsequent equations, we use a
variable in subscript to denote a partial derivative.) The wire is
assumed to extend along $-L \leq x \leq L$, and it is assumed that
$\psi(\pm L,t)=0$. The main conclusions below are also valid for
other boundary conditions such as $\psi_x(\pm L,t)=0$. In order to
concentrate on the main features of the phase transition
mechanism, we take $\sigma=L=1$. This enables us to concentrate on
the key parameters $I$ and $\Gm$. Some TDGL models include a
factor $\zeta_1/\sqrt{1+\zeta_2 |\psi|^2}$ in front of the left
hand side of equation (\ref{g1}). We deal here with the small
$\zeta_2$ limit, but our essential results are valid for finite
positive $\zeta_2$.

\begin{figure}
\begin{center}
\begin{tabular}{c}
\includegraphics[height=7.5cm]{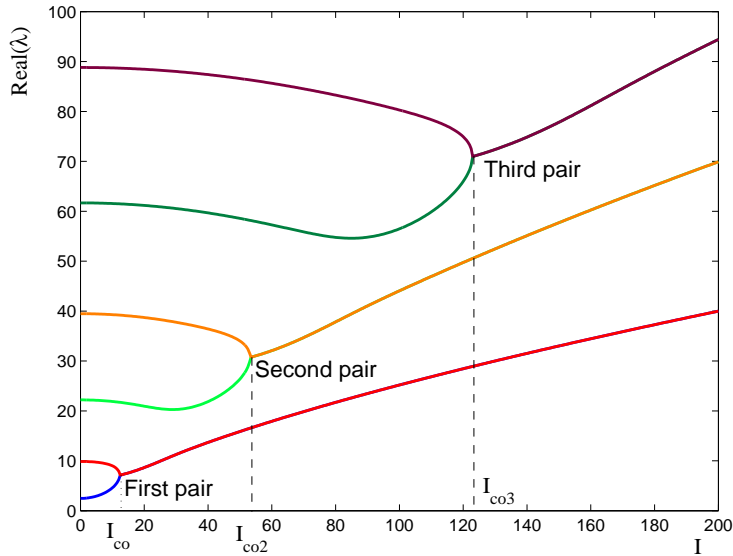}
\end{tabular}
\end{center}
\caption{The real parts of the first 6 eigenvalues of the
PT-symmetric spectral problem (\ref{g7}). } \label{2}
\end{figure}

To understand the phase transition from the normal state to the
superconducting state we linearize the TDGL (\ref{g1}) about the
normal state $\psi=0,\; \phi=-Ix$. Writing
$\psi(x,t)=u(x)e^{(\Gm-\lambda) t}$, we obtain for $u(x)$ the
spectral problem
\begin{equation}
M[u]=u_{xx}+ixIu = -\lambda u,\;\;\;u(\pm 1)=0. \label{g7}
\end{equation}
The spectral problem (\ref{g7}) is called PT-symmetric, since it
is invariant under the joint transformation of $x\rightarrow -x$
(parity) and complex conjugacy (time reversal). The normal state
thus loses its stability when $\Gm > {\rm Real} (\lambda(I))$.
However, since the spectral problem (\ref{g7}) is not
self-adjoint, it is not clear at all that the spectrum is real. On
the other hand the PT symmetry provides some useful information on
the spectrum. Spectral PT-symmetric problems have attracted some
interest in recent years following the numerical observation of
Bender and Boettcher \cite{bebo} that the spectrum of certain
PT-symmetric problems {\em is} real. While ref. \cite{bebo}
considered a problem on the entire line, we deal here with a
problem in a finite interval. When $I=0$ the spectrum is of course
real. The PT-symmetry implies that if $(\lambda,u(x))$ is a
spectral pair with complex $\lambda$, then also
$(\lambda^*,u^*(-x))$ is a spectral pair. Since the spectrum
depends smoothly on $I$ as long as the eigenvalues remain
separated \cite{latr}, a real eigenvalue cannot split
spontaneously into a complex pair. This implies that at least for
small $I$ all eigenvalues are real. However, when the current $I$
is large, the lowest eigenvalues (in absolute value) are shown to
satisfy $\lambda= O(iI)$, namely, to leading order they are purely
imaginary. This implies that eigenvalues indeed collide as $I$
increases. Specifically we find that the first such collision
occurs when the first and second eigenvalues approach each other
and collide at a critical value $I_{co} \approx 12.31$.

At the collision point, the geometric multiplicity of the eigenvalue
is $1$. To find the behavior of the spectrum near $I_{co}$ we set
the current $I$ to be $I=I_{co}+\eps a$. Here $\eps$ is a small
positive number, and we introduce $a$ to determine through its sign
the direction in which we move from $I_{co}$. We then consider an
expansion of the form
\begin{equation}
\lambda=\mu_0+\eps^{1/2} \mu_1 + \eps \mu_2+..., \;\;\;
u=u_0+\eps^{1/2} u_1+\eps u_2 +.... \label{spectrum}
\end{equation}
The nonanalytic nature of the expansion for $\lambda$ is a
consequence of the Jordan form of the spectral problem at the
critical value $I=I_{co}$. The leading order term in
(\ref{spectrum}) is found to be $\mu_0 \approx 0.71$, with an
associated eigenfunction $u_0$ that we normalize by $u_0(0)=1$.
The first order correction $\mu_1$ is conveniently expressed
through the auxiliary function $K(x)$ that solves
\begin{equation}
K_{xx} + ixI_{co} K+\mu_0K=u_0,\; K(\pm 1)=0. \label{keq}
\end{equation}
Writing $u_0=U_r+iU_i$, and defining $a_1=2\int_{-1}^1 xU_rU_i\;
dx$ and $b=\int_{-1}^1 Ku_0\; dx$, one obtains $\mu_1^2=-a a_1/b$.
A numerical integration gives $a_1 \approx 0.29$ and $ b\approx
0.12$. Since $a_1/b >0$, we see that when $a<0$, i.e. when $I$ is
a little smaller than $I_{co}$, there are two real solutions;
these are the first two real eigenvalues just before the
collision. However, for $I$ beyond $I_{co}$, that is for $a>0$,
the single eigenvalue $\mu_0$ splits into a pair of complex
eigenvalues. It can be further shown that $\mu_2$ is a single real
number, i.e. it is the same for both splitting eigenvalues
\cite{rust}. The analysis above shows that the real part of the
leading eigenvalue is not an analytic function of the current at
$I=I_{co}$. In fact, its derivative blows up as $I_{co}$ is
approached from below. On the other hand, the real part of the
first eigenvalue (pair) is a smooth function of $I$ just above
$I_{co}$. This analysis holds for any later collision of real
eigenvalues as well. It is in agreement with the numerical
calculation presented in Figure \ref{2}.

We computed the first few eigenvalues numerically as they increase
past special collision points. Increasing $I$ beyond $I_{co}$, the
first two eigenvalues move as a complex pair according to the
PT-symmetry. The real parts of the first six eigenvalues as a
function of $I$ are plotted in Figure \ref{2}. We see there that
respective pairs of eigenvalues collide at successive critical
values of $I$.

The normal state becomes unstable at that value of $\Gm$ for which
$\Gm-{\rm Real}(\lambda)=0$. For $I<I_{co}$ the first eigenvalue
$\lambda(I)$ is real. When the temperature is sufficiently low,
i.e. when $\Gamma = \lambda(I)$, the normal state loses stability.
Proceeding to high order terms in the bifurcation expansion it is
found that the bifurcation branch that emerges at $\Gamma =
\Gamma_1(I)=\lambda(I)$ is stable for $I<I_k \approx 10.92$. In
this regime, i.e. when $I<I_k$ and $\Gamma > \Gamma(I)$, the
bifurcating solution converges to a stationary solution. By
`stationary' here we mean that writing $\psi=fe^{i\chi}$, the
gauge invariant quantities $f(x,t), \; q(x,t)=\chi_x(x,t)$ and
$\theta(x,t)=\chi_t(x,t)-\phi(x,t)$ converge to stationary
functions $f_0(x),\; q_0(x),\;\theta_0(x)$. Once $I$ crosses the
critical collision value $I_{co}$ and the eigenvalue splits into a
conjugate complex pair, the phase transition temperature is
determined by the condition $\Gm = {\rm
Real}(\lambda(I))=\Gamma_1(I)$. Thus, for $I>I_{co}$ a Hopf
bifurcation occurs and the solution to the full TDGL is periodic.

\begin{figure}
\ghtwo{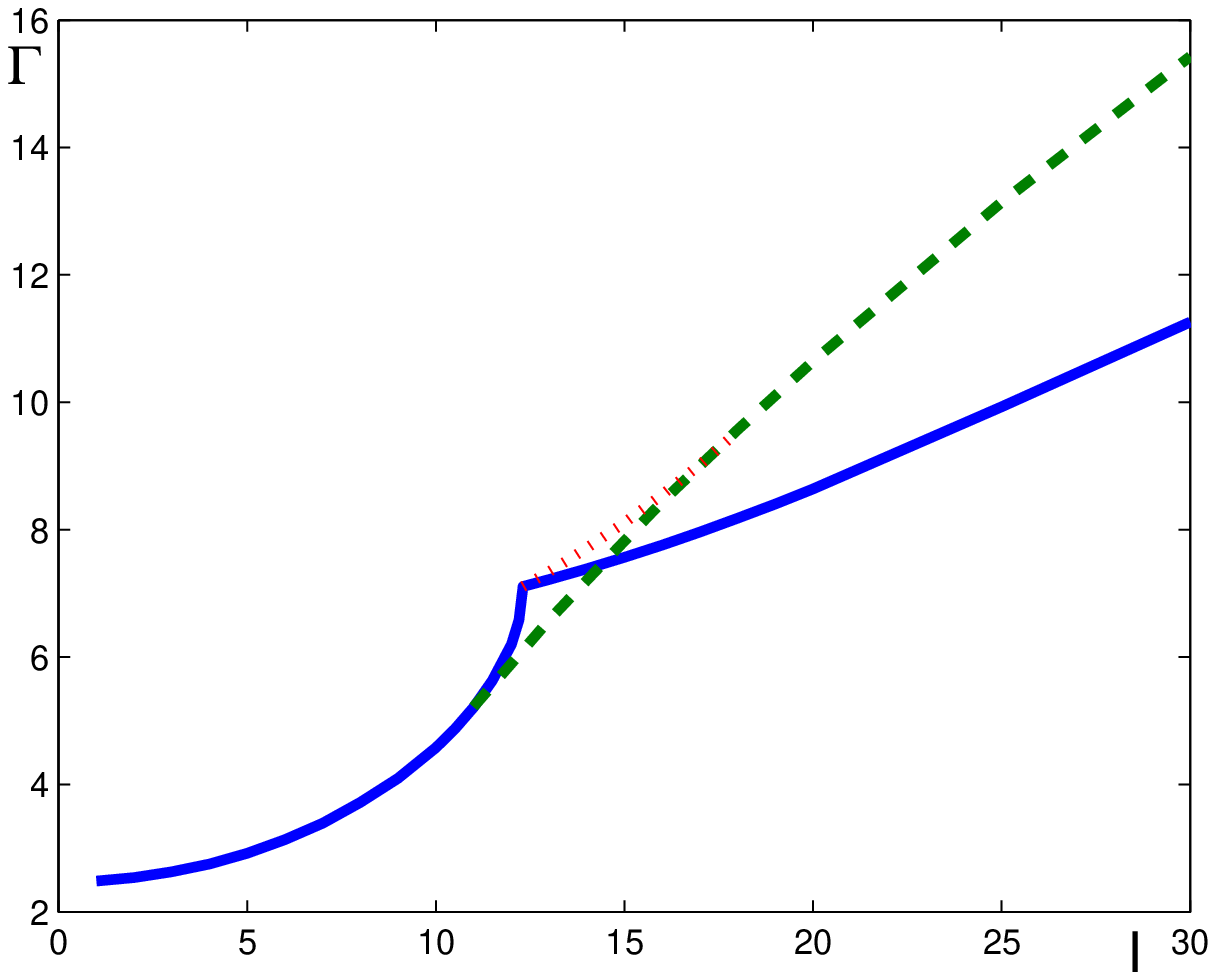}{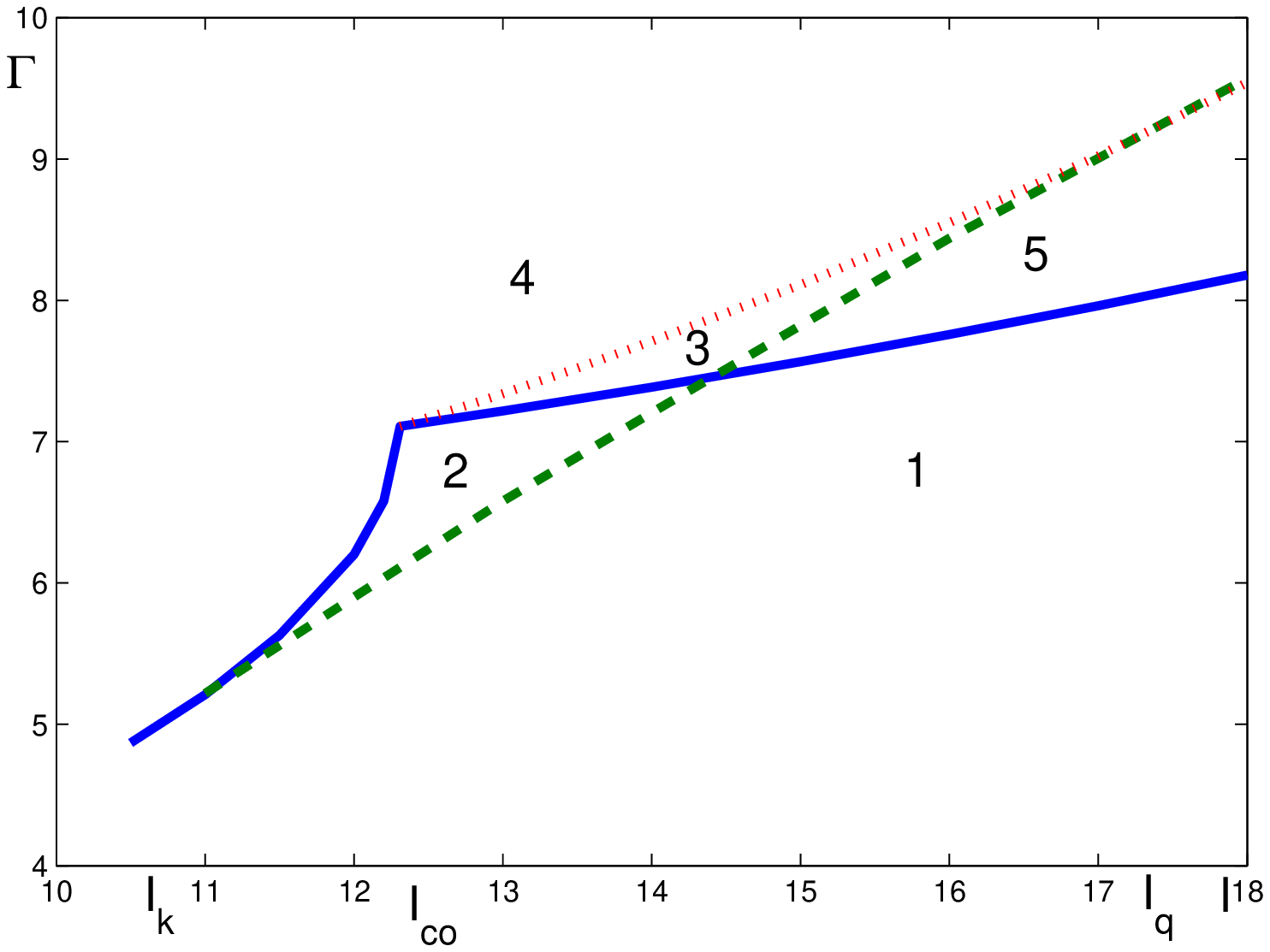} \caption{The phase diagram of the
different stable states in the temperature-current plane. The
parameter $\Gm$ is proportional to $T_c-T$. The curves
$\Gamma_1(I),\;\Gamma_2(I),\;\Gamma_3(I)$ are drawn with solid
line, dashed line and dotted line, respectively. The meaning of
the different curves and regimes is explained in detail in the
text.} \label{1very}
\end{figure}

Consider now a current $I>I_{co}$. When $\Gm$ is below
$\Gamma_1(I)$, the positive real part of the spectrum dominates,
and the normal state is stable. Increasing $\Gm$ with $I$ fixed we
see that when $\Gm=\Gamma_1(I)$ a Hopf bifurcation into a periodic
solution takes place as explained above. In addition to
determining the bifurcation curve $\Gamma = \Gamma(I)$, the
spectral problem (\ref{g7}) can also be used to compute the
bifurcating branch, which is always stable, in the periodic
regime. To see this, fix a current $I$ greater than the critical
value $I_{co}$. Let the ground state of equation (\ref{g7})
consist of the eigenvalue pair $\lm_r \pm i \lm_i$, with
associated eigenfunctions $w_1(x)$ and $w_2(x)$ related by
$w_1(x)=w_2^*(-x)$. We normalize both eigenfunctions by the
condition $w_i(0)=1$. Set the temperature to be slightly below the
critical value determined by $\Gm=\Gamma_1(I)$, by selecting
$\Gm=\lm_r+\eps^2$. Neglecting a short time interval during which
transients decay, the asymptotic solution of the full TDGL
(\ref{g1})-(\ref{g3}) is found to be of the form
\begin{equation}
u(x,t)=\eps A\left(\exp\left(i( \om \eps^2+\lm_i)t\right)w_1(x) +
\exp\left(-i( \om \eps^2+\lm_i\right)t\right)w_2(x)+ O(\eps^3).
\label{asym5}
\end{equation}
The amplitude $A$ and frequency $\om$ are constants that are
determined by the ground state $w_1$ and $w_2$. They are computed
numerically for each current $I$. For instance, when $I=20$ we
found $A=0.921,\; \om =1.8$. One can draw a number of conclusions
from the expression (\ref{asym5}). First, the period of the
oscillations is not exactly the imaginary component of the
eigenvalue, but rather it is has a correction due to the nonlinear
interaction of $w_1$ and $w_2$. Secondly, the solution at $x=0$ is
$u(0,t)\approx 2\eps A\cos\left((\lm_i+\om \eps^2)t\right)$.
Therefore we obtain a phase slip center that is periodic in time
at $x=0$. Ivlev and Kopnin \cite{ivko} made the nice observation
that a PSC can be thought of as a vortex in space-time. In this
sense, the solution structure given in equation (\ref{asym5})
indicates that the PSCs constitute a periodic placement of
degree-one space-time vortices with period
$P=\pi/(\lambda_i+\omega \eps^2)$. The curve $\Gamma_1(I)$ is
depicted by the solid line in Figure \ref{1very}.

So far we have concentrated on the smooth bifurcation of the
normal state into a periodic state or into a stationary state. It
turns out, though, that there are regions in the parameter plane
where two metastable states coexist. The transition between them
is nonsmooth, and therefore it is associated with hysteresis. We
already pointed out above that for $I_{co}>I>I_k$ the normal state
bifurcates into an unstable branch. This hints that the phase
transition there is nonsmooth. Indeed, we identified another curve
in the phase plane, that we call $\Gamma_2(I)$, above which the
stationary state is stable. The curve $\Gamma_2(I)$ is depicted as
a dashed line in Figure \ref{1very}.

To understand the loss of stability of the periodic state, we
recall that the Hopf bifurcation that led to it was triggered by
the normal current contribution to the potential term $i \phi
\psi$ in equation (\ref{g1}). Near the transition curve $\Gm =
\Gamma_1(I)$ the magnitude $|\psi|$ of the order parameter is
still small, and essentially the entire current is normal. As the
temperature is lowered (i.e. $\Gm$ increases), $|\psi|$ grows and
so does the supercurrent, implying via (\ref{g3}) that the normal
current decreases. This effectively returns the system to the
small $I$ regime where the bifurcation to a steady state is
favored. We thus obtain a third bifurcation curve $\Gamma_3(I)$
where the periodic state loses it stability.

At this point we make reference to Figure \ref{1very} and consider
the different regimes in the $(I,\Gamma)$ plane. The solid curve
provides the critical temperature $\Gm=\Gamma_1(I)$ along which
the normal state loses its stability. A stable stationary state
exists above the dashed line that represents a second curve
$\Gamma_2(I)$. For $I<I_{k}$ the normal state bifurcates into a
stable stationary superconducting state. For $I>I_{co}$, on the
other hand, the normal state bifurcates into a state that exhibits
time-periodic oscillations. When $I>I_{co}$, and the temperature
is further lowered ($\Gm$ is increased), the periodic state loses
its stability at a third critical temperature $\Gm=\Gamma_3(I)$
represented by the dotted line in the figure. The curves
$\Gamma_2(I)$ and $\Gamma_3(I)$ intersect at $I_q$. For $I>I_q$,
the curves $\Gamma_2(I)$ and $\Gamma_3(I)$ coalesce. The frame on
the left depicts the bifurcation curves over a large $(I,\Gamma)$
area, while the frame on the right concentrates on the interesting
area near the point $(I_{co},\Gamma_1(I_{co}))$, where
$\Gamma_1(I_{co}) \approx 7.11$. The parameter plane is
partitioned into 5 domains. In domains 1,4 and 5 there is a single
stable state - the normal state in region 1, a stationary state in
region 4 and a periodic state in region 5. In region 2 there are
two metastable states - normal and stationary, while in region 3 a
stationary and a periodic state are both metastable.

\begin{figure}
\begin{center}
\begin{tabular}{c}
\includegraphics[height=6cm]{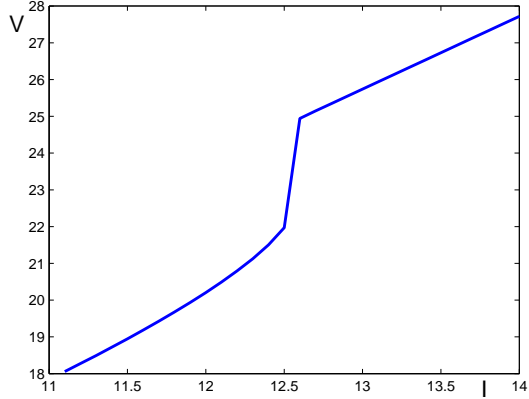}
\end{tabular}
\end{center}
\caption{The I-V curve for $\Gamma=6.3$ and $I$ increasing. Notice
that, although the periodic state does not exist for such
temperature, and thus there is no PSC here, the I-V curve does
exhibit a jump discontinuity at $I \approx 12.57$ }
\label{IVg63r01+2}
\end{figure}

We proceed to draw two further conclusions related to the
bifurcation diagram. It is useful to do so in the context of I-V
curves. These curves are measured or computed for a fixed
$\Gamma$, while the current $I$ is raised or lowered
adiabatically. When this process cuts through the metastable
regions 2 and 3 in Figure \ref{1very} a hysteresis is expected in
the I-V curve. While such a hysteresis was predicted a long time
ago, we point out that it is not always observed experimentally
\cite{vodo}. As can be seen in Figure \ref{1very}, the metastable
regions are quite small, and therefore it requires careful tuning
to pass through them. Another comment relates to the formation of
PSCs. These points in space-time where $|\psi|$ vanishes are often
associated in the literature with jump discontinuities in the I-V
curve. However, this identification works only in one way, and not
all such jumps imply the presence of a PSC. For instance, we
depict in Figure \ref{IVg63r01+2} the I-V curve for $\Gamma=6.3$
and $I$ slowly increasing. For this $\Gamma$ one never crosses an
area in the parameter plane where the periodic state is stable,
and therefore there is no PSC. Nonetheless, the I-V curve exhibits
a clear discontinuity at about $I \approx 12.57$. The actual rule
for lack of smoothness in I-V curves is that a jump discontinuity
indicates a nonsmooth phase transition, while a discontinuity in
the derivative indicates a continuous phase transition.

To summarize, using a combination of asymptotic expansions,
spectral analysis and canonical numerical computation applied to
the time-dependent Ginzburg-Landau model, we have presented a full
analysis of the behavior of a one-dimensional superconducting wire
exposed to an applied electric current. In particular, retaining
temperature and applied current as parameters, we have decomposed
this two-dimensional parameter space into regions of stability of
a normal, stationary and oscillatory state. The collision of real
eigenvalues and the consequent emergence of complex spectrum in
the associated linearized problem provides the explanation for the
Hopf bifurcation leading to the appearance of the oscillatory
state and the associated phase slip centers. From the theoretical
standpoint, it also reveals a physically significant setting where
PT-symmetry does {\em not} lead to reality of the spectrum, in
contrast to its common role \cite{bebo}, \cite{cdv}. The boundary
of the basin of attraction of the normal state has been given
precisely in terms of the real part of the leading eigenvalue in
this linearized problem. The boundary between the basins of
attraction of the oscillatory and stationary states has been
calculated near the triple point using asymptotics, and has been
computed numerically beyond this. In so doing, we have identified
small regions in the parameter space where hysteresis should be
anticipated. Finally the asymptotic structure of the periodic
solution bifurcating off of the normal state has been developed
for I above the first collision value $I_{co}$ and for $\Gamma$
just above the real part of the first eigenvalue. This expansion
reveals the period of the oscillations and location of PSCs along
the $x$-axis

\end{document}

%% file: graph.tex
\input epsf.sty

\def\goneL#1{\vskip 8pt
\centerline{\epsfysize=8.0truein{\epsfbox{ #1 }} \kern0pt }
\vskip 8pt}

\def\gonel#1{\vskip 8pt
\centerline{\epsfysize=4.5truein{\epsfbox{ #1 }} \kern0pt }
\vskip 8pt}

\def\gone#1{\vskip 8pt
\centerline{\epsfysize=3.9truein{\epsfbox{ #1 }} \kern0pt }
\vskip 8pt}

\def\gonem#1{\vskip 8pt
\centerline{\epsfysize=3.2truein{\epsfbox{ #1 }} \kern0pt }
\vskip 8pt}

\def\gones#1{\vskip 8pt
\centerline{\epsfysize=2.5truein{\epsfbox{ #1 }} \kern0pt }
\vskip 8pt}

\def\gonemini#1{\vskip 8pt
\centerline{\epsfysize=1.8truein{\epsfbox{ #1 }} \kern0pt }
\vskip 8pt}

\def\ghtwo#1#2{\vskip 8pt
\centerline{
\epsfysize=2.5truein{\epsfbox{ #1 }}
\epsfysize=2.5truein{\epsfbox{ #2 }} \kern0pt }
\vskip 8pt}

\def\gvtwo#1#2{\vskip 8pt
\centerline{
\epsfysize=2truein{\epsfbox{ #1 }} \kern0pt}
\centerline{
\epsfysize=2truein{\epsfbox{ #2 }} \kern0pt }
\vskip 8pt}

\def\gthree#1#2#3{\vskip 8pt
\centerline{
\epsfysize=2.5truein{\epsfbox{ #1 }}\kern0pt}
\vskip 0.1cm
\centerline{
\epsfysize=2.5truein{\epsfbox{ #2 }}
\epsfysize=2.5truein{\epsfbox{ #3 }} \kern0pt }
\vskip 8pt}

\def\gfour#1#2#3#4{\vskip 8pt
\centerline{
\epsfysize=2.0truein{\epsfbox{ #1 }}
\epsfysize=2.0truein{\epsfbox{ #2 }} \kern0pt }
\centerline{
\epsfysize=2.0truein{\epsfbox{ #3 }}
\epsfysize=2.0truein{\epsfbox{ #4 }} \kern0pt }
\vskip 8pt}